\documentstyle{europhys}

\def\And{{\rm and\ }}

\newif\ifboo \boofalse

\def\Review#1{\boofalse{\it #1},}
\def\Name#1{{\sc #1},}
\def\Vol#1{\ifboo Vol. {\bf #1}\else{\bf #1}\fi}
\def\Year#1{\ifboo #1\else(#1)\fi}

\def\Page#1{\ifboo {\rm p. #1}\else{\rm #1}\fi}

\input epsf.sty
\begin{document}
\euro{}{}{}{}
\Date{}
\shorttitle{G. BENENTI {\it et al.}: Intermediate low temperature Coulomb 
$2d$ phase}  
\title{Signatures of an intermediate $2d$ Coulomb phase 
at low temperatures}  
\author{Giuliano Benenti, Xavier Waintal \And Jean-Louis Pichard}
\institute{
CEA, Service de Physique de l'Etat Condens\'e, \\
Centre d'Etudes de Saclay, F-91191 Gif-sur-Yvette Cedex, France   
}
\rec{}{}
\pacs{
\Pacs{71}{30+h} {Metal-insulator transitions and other electronic transitions}
\Pacs{71}{27+a} {Strongly correlated electron systems} 
\Pacs{72}{15Rn} {Quantum localization}
}
\maketitle
\begin{abstract}
 The study of the ground state of spinless fermions in $2d$ 
disordered clusters (Phys. Rev. Lett. {\bf 83}, 1826 (1999)) has 
suggested the existence of a new quantum phase for intermediate 
Coulomb energy to kinetic energy ratios $r_s$. Exact diagonalization 
of the same small clusters show that its low energy excitations 
(quantum ergodicity above a few ``hexatic'' excitations characterized 
by oriented currents) significantly differ from those occuring in the 
Fermi glass (weak $r_s$) and in the pinned Wigner crystal (large $r_s$). 
The ``hexatic'' excitations vanish for temperatures of order of the 
Fermi temperature. 
\end{abstract}

%
%

 An insulator-metal transition (IMT) has been observed when 
the carrier density $n_s$ is increased in dilute two dimensional 
($2d$) gases of electrons \cite{exp1,kravklap} and holes \cite{yoon} having in 
common very low Fermi temperatures ($T_F \approx 1$ K). The critical 
Coulomb energy to Fermi energy ratios $r_s \propto 1/\sqrt{n_s}$  
are typically around $10-35$ at the IMT, dropping as a function of the 
elastic scattering rate $1/\tau$ from $r_s\approx 35$ for a very 
clean heterostructure \cite{yoon} to a value of order $10$ 
when $\tau \leq 10^{-11} s$. This coincides with the critical values 
$r_s^W$ found in numerical simulations for the melting of the Wigner 
crystal, in the clean limit ($r_s^W \approx 37$) and in the disordered 
case ($r_s^W \approx 10$, see Refs. \cite{yoon,lettre1}). In the 
insulating phase ($r_s > r_s^W$), $I-V$ characteristics \cite{yoon,pudalov} 
consistent with the existence of a pinned crystal have been observed, 
accompanied by a sharp drop of the compressibility \cite{comp}. 
Therefore, it has been suggested \cite{yoon,lettre1,pudalov} that 
the insulating character of the phase is due to the pinning of a 
Wigner crystal, and not to Anderson localization. At higher densities 
($r_s \leq r_s^F \approx 6$), the kinetic terms become more important, giving 
rise to a re-entry where the metallic behavior disappears. In this limit, 
the insulating behavior predicted by the scaling theory of localization 
is recovered \cite{hamilton} (Fermi glass). The metallic behavior 
characterizing 
intermediate ratios $r_s^F < r_s < r_s^W$  persists \cite{kravklap} down to 
$ T \rightarrow 35$ mK ($T/T_F \rightarrow 10^{-2}$), but vanishes 
\cite{exp1,yoon} when the temperature $T$ exceeds the Fermi temperature 
$T_F$. Eventually, the IMT occurs when the elastic mean free path $l$ and 
the Fermi wave vector $k_F$ are such that $k_Fl \approx 1$.  In summary, 
the interaction, disorder and kinetic effects are {\it a priori} relevant 
and non pertubative in those dilute gases and the observation of the IMT 
in different materials leads us to suspect the existence of a rather general 
mechanism.

 Although a generalization up to $r_s \rightarrow 10$ of the 
non-interacting theory where the disorder depends on temperature 
has been proposed to partly explain \cite{altshuler} the metallic 
behavior, we numerically explore another alternative: Would the IMT be one 
of the signatures of a new quantum phase (Coulomb metal) driven by 
the interplay between kinetic energy and Coulomb repulsion 
in a random substrate? A first numerical study \cite{lettre1} was 
focused on the ground state of spinless fermions with Coulomb repulsion 
in disordered clusters forming a $2d$ torus enclosing an Aharonov-Bohm 
flux $\phi$ along the longitudinal direction. The topology of the 
pattern of the driven currents provides the signature 
of a new intermediate quantum regime. At weak $r_s$ (Fermi glass) one has a 
$2d$ pattern where the local 
current flows in every direction. At intermediate $r_s$ (new regime), 
the pattern becomes essentially $1d$ (no transverse currents and enhanced 
$1d$ longitudinal flows enclosing the flux, see also Ref.\cite{ab}). 
At large $r_s$ (pinned 
Wigner crystal), the persisting $1d$ flows disappear and charge 
crystallization sets in. This suggests the possible existence of a 
new quantum phase characterized by an orientational order of the 
persistent currents without charge crystallization. This phase could 
be a quantum analog in a disordered system of the hexatic phase 
supposed to occur at the melting of the {\it classical} Wigner 
crystal~\cite{halperin} above the Kosterlitz-Thouless temperature. 
Characteristic values $r_s^F$ and $r_s^W$ delimiting 
the new intermediate regime were obtained in agreement with the 
observed re-entry ($r_s^F \approx 4$) and IMT ($r_s^W \approx 10$).

In this Letter, we study the low energy excitations of the same
clusters which were studied in Ref. \cite{lettre1}, and their 
statistics when the microscopic configurations of the random substrate 
are changed. We find that the characteristic low energy excitations 
of the new intermediate regime disappear for 
excitation energies $\epsilon$ of order of the Fermi energy $\epsilon_F$.
Inside the intermediate regime, the low energy excitations ($\epsilon 
\leq \epsilon_F$) are characterized by a non random orientation of the 
local currents and deviations from Wigner-Dyson (W-D) spectral statistics. 
In analogy with the classical Coulomb problem, we refer to these oriented 
non chaotic excitations as ``hexatic''. For excitation energies 
$\epsilon \geq \epsilon_F$, the local currents become 
randomly oriented and the levels obey W-D statistics (quantum ergodicity). 
In the Fermi glass and in the Wigner crystal, ergodicity is expected 
for energies much larger than $\epsilon_F$ and the levels remain 
uncorrelated for $\epsilon \approx \epsilon_F$. Quantum ergodicity with 
correlated energy levels for $\epsilon \geq \epsilon_F$ is then another 
characteristic of the intermediate regime, in addition to ``hexatic'' 
excitations for $\epsilon \leq \epsilon_F$. Moreover, the sensitivity 
of the levels under a change of the threaded flux has a remarkable 
temperature dependence. When $T \rightarrow T_F$, the enhancement 
observed in the intermediate regime is suppressed while the weak 
responses of the Fermi and Wigner limits are enhanced. Eventually, 
assuming that this intermediate regime observed in small clusters 
is the signature of a new quantum phase, we propose the local current 
$\vec{j}$ driven by a flux $\phi$ as a possible order 
parameter for the successive quantum transitions. The first transition 
would be associated to the angle $\theta$ of $\vec{j}$ with the direction 
enclosing the flux and the second one to its amplitude $j$.

%
%

We consider a disordered square lattice with  $N=4$ spinless fermions 
on $L^2=36$ sites. The Hamiltonian 
reads 
\begin{equation} 
\label{hamiltonian} 
H=-t\sum_{<i,j>} c^{\dagger}_i c_j +  
\sum_i v_i n_i  + U \sum_{i\neq j} \frac{n_i n_j } {2r_{ij}},
\end{equation} 
where $c^{\dagger}_i$ ($c_i$) creates (destroys) an electron on 
site $i=(i_x,i_y)$, the hopping term $t$ between nearest neighbours 
characterizes the kinetic energy, the site potentials $v_i$ are  
taken from a box distribution of width $W$, $n_i=c^{\dagger}_i c_i$ 
is the occupation number at the site $i$ and $U$ measures the 
strength of the Coulomb repulsion. $r_{ij}$ is the inter-particle 
distance for a $2d$ torus. In our units, the factor $r_s
= U (2t\sqrt{\pi n_e})^{-1}$ at a filling factor $n_e=N/L^2$. 
The Fermi energy is defined by $\epsilon_F=4 \pi n_e t$ and 
a Fermi golden rule approximation gives $k_F l\approx 192\pi n_e (t/W)^2$. 
We consider $W/t=5$ corresponding to $k_F l=2.7$. The boundary 
conditions (BCs) are always taken periodic in the transverse $y$-direction, 
and such that the system encloses an Aharonov-Bohm flux $\phi=\pi/2$ in 
the longitudinal $x$-direction ($\phi=\pi$ corresponds to anti-periodic BCs). 
The Hamiltonian is diagonalized using the Lanczos 
algorithm for a statistical ensemble of $10^3$ clusters. The first 
$20$ energy levels $E_n$ ($n=0, 1,2, \ldots$) are considered for values 
of $r_s$ chosen inside the three phases ($r_s=0.8$ for the Fermi glass, 
$r_s=6.3$ for the conjectured intermediate phase and $r_s = 30$ for the 
pinned Wigner crystal). 

%
%

 We begin the presentation of our results by discussing the spectral 
fluctuations. For the one body spectra, localized 
wavefunctions yield uncorrelated spectra with Poisson statistics, 
characterized by a distribution $P(s)$  of the energy spacings between 
successive levels going to $P_P(s)=\exp(-s)$ when $L \gg L_1$. Delocalized 
wavefunctions yield correlated spectra with W-D statistics. With 
time-reversal symmetry ($\phi=0$), $P(s) \rightarrow 
P_W^O(s)=(\pi s/2)\exp(-\pi s^2/4)$ as $L$ increases, and without 
time-reversal symmetry ($\phi=\pi/2$), $P(s) \rightarrow 
P_W^U(s)=(32 s^2/\pi^2)\exp(-4 s^2/\pi)$. The statistical properties of 
many body spectra are more complex. Without interaction, the levels 
are uncorrelated, and a sufficient two body interaction may restore 
W-D statistics. This is at the basis of the success of random 
matrix theory to describe complex nuclei, where low energy quasi-particle or 
collective excitations are followed by chaotic high energy excitations. 
The low energy modes being close to integrability display Poisson 
statistics as non-generic integrable models, whereas in the absence 
of integrals of motion quantum ergodicity and W-D statistics 
set in \cite{montambaux}. The question is to know at what energy 
threshold the Poisson-Wigner crossover takes place. An answer was 
proposed in Ref. \cite{wpi,js} based on the breakdown of a perturbation 
theory valid for weak $U$ (the matrix element of a Slater determinant 
to the ``first generation'' of determinants directly coupled to it by 
interaction is of order of the level spacing of the latter determinants). 
This question was recently addressed \cite{berkovits,song} 
for the Hamiltonian which we consider, showing as expected quantum 
ergocity at sufficient excitation energies, but failing to check the 
proposed criterion (the estimation of the matrix elements deserving a 
particular study). Another question is to know the factors $r_s$ delimiting 
the regime of quantum ergodicity for a given excitation energy $\epsilon$. 
The qualitative answer can be guessed from a simpler problem~\cite{tip2,sp1} 
(two interacting particles in $1d$) where a  duality $t/U \leftrightarrow 
U/t$ implies that a maximum spectral rigidity occurs for $U \approx t$. 
This leads us to a general picture where Fermi glass and Wigner crystal 
should be associated to integrability and Poisson statistics, in contrast 
to an intermediate regime displaying quantum ergodicity at low excitation 
energies. 

\begin{figure}
\centerline{  
\epsfxsize=14cm
\epsfbox{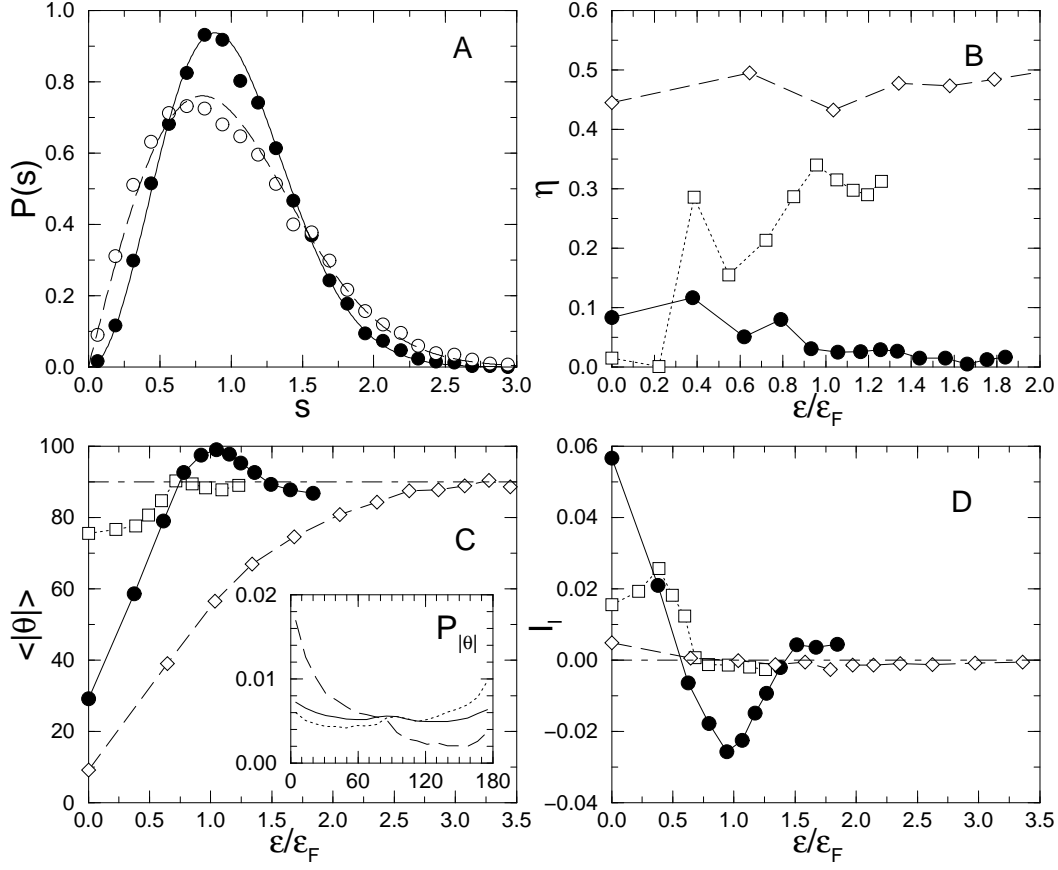}
}
\caption{ 
A: Spacing distribution $P(s)$ for $r_s=6.3$ 
when $\phi=0$ (empty circles) and $\phi=\pi/2$ (full circles) 
at excitation energies $1.4  < \epsilon/\epsilon_F < 1.9$, 
compared to $P_O^W(s)$ (dashed line) and 
$P_U^W(s)$ (continuous line). B: parameter $\eta$ as a function of 
$\epsilon/\epsilon_F$ for $\phi=\pi/2$. $r_s=0.8$ 
(Fermi glass, squares), $r_s=6.3$ (intermediate regime, full circles) 
and $r_s=30$ (Wigner crystal, diamonds). C (Ensemble average 
$<|\theta|>$) and D (typical total longitudinal current $I_l$) 
as a function of $\epsilon/\epsilon_F$ (same cases and symbols than 
in B). Inset of C: angular distribution $P_{|\theta|}$ in the intermediate 
regime ($r_s=6.3$) for $\epsilon/\epsilon_F=0.4$ (dashed line), $1$ 
(dotted line) and $1.4 - 1.9$ (solid line), respectively.
}\label{fig1}
\end{figure}

We study the distribution of the normalized $N$-body energy spacings 
$s_n=(E_{n+1}-E_n)/<E_{n+1}-E_n>$ (the brackets denote ensemble 
average).  In Fig. \ref{fig1} A, one can see that 
$P(s)$ is given by the Wigner surmises, with an orthogonal-unitary 
crossover when $\phi =0 \rightarrow \pi/2$, in the intermediate regime 
for $ 1.4 < \epsilon/\epsilon_F < 1.9$ ($n=10, \ldots ,20$). We define 
a parameter 
\begin{equation}
\eta=\frac{\hbox{var}(P(s))-\hbox{var}(P_W^U(s)))}   
{\hbox{var}(P_P(s))-\hbox{var}(P_W^U(s)))},
\end{equation}
($\hbox{var}(P(s))$ denoting the variance of $P(s)$). 
$\eta=1$ when $P(s)=P_P(s)$ and $\eta=0$ when $P(s)=P_W^U(s)$. 
Taking $\phi=\pi/2$, the variation of $\eta$ as a function of 
$\epsilon/\epsilon_F$ is given in Fig. \ref{fig1} B in the three phases. 
For the Fermi glass with $r_s=0.8$, the two first spacings 
display  $\eta$ values compatible with the unitary ensemble, 
while the following spacings become more and more Poissonian. 
This is what is expected for $r_s = 0$, the two first 
excitations being single-electron transitions with W-D  
statistics for $L<L_1$, while the following excitations being the sum 
of more than one single-electron excitation, give uncorrelated levels. 
For a small $r_s$, this still remains true, the system being composed 
of weakly interacting quasi-particles. For the Wigner crystal ($r_s=30$), 
$\eta$ does not depend on $\epsilon$ up to $3\epsilon_F$ and keeps 
an intermediate value, $\eta=1$ being expected in the thermodynamic 
limit. For the intermediate regime ($r_s=6.3$) one has quantum ergodicity 
and W-D statistics above a (low) excitation energy $\approx \epsilon_F$, 
but the first excitations exhibit deviations from W-D statistics, 
indicating a break-down of a description in terms of weakly interacting 
quasi-particles and suggesting possible non chaotic collective excitations.  

%
%

 The low energy excitations are characterized by a particular 
structure of the corresponding persistent currents when a 
flux $\phi$ is applied. The current $I^{(n)}$ of the $n^{\rm th}$ 
many-body level has a total longitudinal component given by 
\begin{equation} 
I_l^{(n)}(\overline{\phi})=-\left.\frac{\partial E_n}{\partial\phi} 
\right|_{\phi=\overline{\phi}}=\frac{\sum_{i}j_{i,l}^{(n)}}{L}.
\end{equation} 
The local current $j_{i,l}^{(n)}$ flowing at the site $i$ in the 
longitudinal direction is defined as 
\begin{equation} 
j_{i,l}^{(n)}=2\hbox{Im}\langle\Psi_n|c_{i_x+1,i_y}^\dagger
c_{i_x,i_y}\exp(i\phi / L)|\Psi_n\rangle, 
\end{equation}
where $\Psi_n$ $(n=0,1,2,...)$ is the many-body wavefunction 
of energy $E_n$. One defines the local transverse current 
$j_{i,t}^{(n)}$ by a similar expression. The local current of the 
$n^{\rm th}$ level flowing at the site $i$ is a two-component vector 
$\vec{j}^{(n)}(i)=(j_{i,l}^{(n)},j_{i,t}^{(n)})$
characterized by its angle 
$\theta_i^{(n)}=\arctan(j_{i,l}^{(n)}/j_{i,t}^{(n)})$ 
and its absolute value $j_i^{(n)}=|\vec{j}^{(n)}(i)|$.
 
  The inset of Fig. \ref{fig1} C gives the distribution 
$P_{|\theta|}$ of $|\theta|$ in the intermediate regime for 
increasing excitation energies ($P_{\theta}$ is symmetric 
when $\theta \rightarrow -\theta$).  For $\epsilon/\epsilon_F  \geq 
1.4$ where W-D rigidity sets in, $P_{|\theta|}$ is uniform, 
the flow is randomly oriented. For $ \epsilon/\epsilon_F \leq 1.4$ 
where deviations from W-D statistics can be seen, 
$P_{|\theta|}$ is non uniform, the flow being mainly paramagnetic 
at $\epsilon/\epsilon_F \approx 0.4$ and diamagnetic at 
$\epsilon/\epsilon_F \approx 1$. The average of $|\theta|$ is 
shown in Fig.\ref{fig1} C: values $< \pi/2$ or $> \pi/2$ correspond  
to paramagnetic or diamagnetic responses respectively. In the Fermi glass, 
the flow is weakly paramagnetic and becomes randomly oriented for 
$\epsilon > \epsilon_F$. In the pinned Wigner crystal, a strongly 
paramagnetic response becomes randomly oriented at $\epsilon > 3 \epsilon_F$. 
The intermediate regime has a very specific behavior, the flow 
being mainly paramagnetic ($\epsilon/\epsilon_F <0.7$) then diamagnetic 
($\epsilon/\epsilon_F \approx 1 $), before being randomly oriented at 
$\epsilon/\epsilon_F \geq 1.4$ when quantum ergodicity sets in. The 
amplitudes of the longitudinal paramagnetic (probability $c_p$) 
and diamagnetic (probability $c_d$) total current $I_l (E_n \approx 
\epsilon)$ have log-normal distribution. The typical value 
$I_l= c_p \exp <\ln |I_l^p|> - c_d \exp <\ln |I_l^d|>$ is given 
in Fig. \ref{fig1} D. Below the ergodicity threshold of the 
intermediate regime, $I_l$ 
is non random, being paramagnetic then diamagnetic in contrast to the 
Fermi and Wigner limits. The sign of $I_l$ can be 
explained \cite{weinsel} in the Wigner limit, where the 
$\phi$ dependent parts of the energies $E_n$ can be expanded in power 
of $t/U$. The leading contribution comes from the shortest $1d$ paths 
enclosing $\phi$ and the sign is given by Leggett's theorem (see 
Ref.\cite{lettre1}) for $1d$ spinless fermions. The sign depends on 
the sample geometry and the parity of $N$  when 
$r_s \rightarrow \infty$, independently of the microscopic disorder, 
but remains the same as far as $r_s \geq r_s^F$, becoming sample 
dependent in the Fermi glass only.

\begin{figure}
\centerline{
\epsfxsize=14cm
\epsfbox{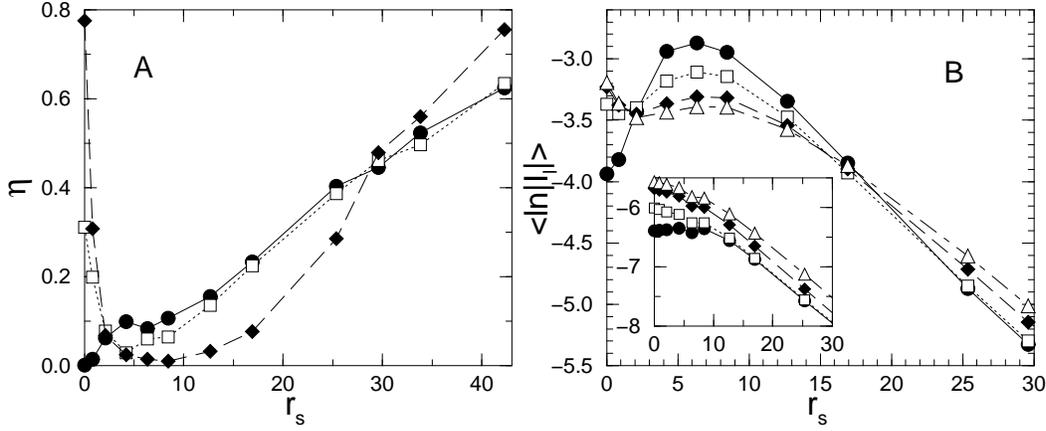}
}
\caption{ $W/t=5$ and $\phi=\pi/2$. A: Parameter $\eta$ as a 
function of $r_s$ for the first excitation $s_0$ (circles), 
$s_2-s_4$ (squares) and $s_{10}-s_{20}$ (diamonds). 
B: Typical longitudinal current for 
temperature in units of the Fermi temperature $T/T_F=0$ (circles), 
$0.4$  (squares), $0.7$ (diamonds), $1.1$ (triangles up). Inset: 
the same at $W/t=15$.
}
\label{fig2}
\end{figure}
 
   Additional informations on the spectral statistics are given in Fig. 
\ref{fig2} A ($\eta$ as a function of $r_s$ for levels of 
increasing excitation energies). There are two values of $r_s$ 
($\approx 2$ and $\approx 30$) where $P(s)$ is invariant when 
$\epsilon/\epsilon_F \leq 3$. Between those two values, the spectrum 
becomes more rigid as one increases $\epsilon$, while the opposite 
behavior is observed otherwise, providing another signature of the 
intermediate regime. Assuming a Gibbs-Boltzmann population of the 
many-body levels, we show in Fig. \ref{fig2} B the variation of 
\begin{equation}
<\ln |I_l (kT)|> =  \left< \frac{\sum_n \ln|I_l^{n}| \exp (- (E_n-E_0)/kT)}
{\sum_n \exp (-(E_n-E_0)/kT)} \right>
\end{equation}
as a function of $r_s$ at different temperatures. Another signature 
of the intermediate regime is obtained, the enhancement of the 
above expression disappearing as $T \rightarrow T_F$, while there is 
a weak increase 
in the Fermi and Wigner limit. We point out that the metallic 
(insulating) phase is identified in the transport measurements 
from a decrease (increase) of the conductivity when $T$ increases 
and that the low temperature metallic behaviour disappears for 
$T\approx T_F$. In the inset, one can see that enhancement 
and opposite temperature dependence disappear for a larger disorder 
($W/t=15$), indicating that the ratio $W/t$ is another relevant parameter, 
a metallic behavior being not expected \cite{hf} for too strong disorders.   

\begin{figure}
\centerline{ 
\epsfxsize=14cm
\epsfbox{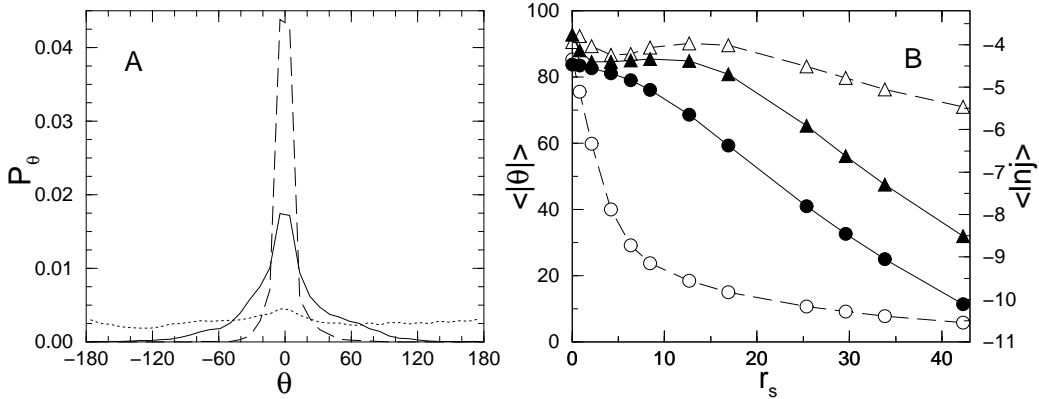}
}
\caption{
$\phi=\pi/2$. A: distribution $P_{\theta}$ of the ground state 
local current angles for $r_s=0$ (dotted line), 
$r_s=6.3$ (full line) and $r_s=42$ (dashed line). B: 
$<|\theta|>$ (open symbols, left scale) and amplitudes $<\ln j>$ 
(filled symbols, right scale) as a function of $r_s$ for the 
ground state (circles) and for $1.4 <\epsilon/\epsilon_F<1.9$ 
(triangles). 
}
\label{fig3}
\end{figure}

 The distribution $P_{\theta}$ calculated for the ground state 
( Fig. \ref{fig3} A) shows how the local angles undergo a 
transition from a nearly uniform angular distribution (glass of 
currents randomly directed in the plane) towards a distribution 
strongly peaked in the longitudinal direction (ordered flow of 
currents with a $1d$ topology). Fig. \ref{fig3} B gives 
the average angle $<|\theta|>$ and the local current amplitude 
$<\ln j>$ when $r_s$ increases, both for the ground state and 
for $\epsilon > \epsilon_F$: The pattern of driven currents 
becomes ordered for $r_s$ values smaller than those at 
which $j$ is suppressed due to charge crystallization. 

 In summary, the local currents and the level statistics 
characterizing the low energy excitations confirm the existence 
of an intermediate regime in small clusters. The main issue is to 
know if this intermediate regime gives rise to a new quantum phase 
in the thermodynamic limit. A finite size scaling (FSS) study using 
exact diagonalization (larger $L$ at the same density $n_e$) is beyond 
the ability of the most powerful available computers. One alternative 
could be to use variational trial wavefunctions justified for small 
and large $r_s$ and to compare their relative energies using Monte Carlo 
method \cite{montecarlo}. However, such approaches depend on the 
taken trial wavefunctions and can easily miss complex behaviors as the 
ones behind the beautiful FQHE physics, where the exact numerical 
solution \cite{laughlin} of the three particle problem have been 
instructive. To study larger sizes, we have also developed approximate 
methods where the huge Hamiltonian is truncated in restricted sub-spaces 
using one body states \cite{lettre2} or Hartree-Fock orbitals \cite{hf}. 
From those studies valid for small and large $r_s$, but which become 
unreliable for intermediary values of $r_s$, we know that the obtained 
precisions do not allow us to calculate very small quantities (as the local 
currents presented here). Numerical studies of those Coulomb systems are 
limited to exact diagonalizations of small clusters combined with approximate 
FSS approaches. For instance, a FSS study \cite{lettre2} of the length $\xi$ 
characterizing the change of the ground state density after a local 
perturbation of the substrate has revealed the existence of a divergence 
when $r_s \rightarrow r_s^F$, giving the signature of a real quantum 
transition. In this case, we propose the 
local current $\vec{j}$ as a possible order parameter driving the transitions 
from the intermediate phase to both the Fermi glass (via 
its angle $\theta$) and the pinned Wigner crystal (via its amplitude $j$). 
The melting of the Wigner crystal leads to an intermediate ``hexatic'' 
phase which has no translational order but still has orientational order  
(alignment of the local currents $\vec{j}$). Shear modulus measurement 
\cite{williams} would be very useful, together with transport and 
compressibility measurements, to clarify these issues. We note that 
the disorder strength $W$ should play a role which has not been fully 
explored (the study is mainly centered on $k_Fl \approx 2.7$). The role of 
the spin degrees of freedom will be discussed in details in a following 
study \cite{sp2} which does not lead us to fundamentally reconsider our 
conclusions.


%

\end{document}